\documentclass{article}

\usepackage{arxiv}

\usepackage[utf8]{inputenc} 
\usepackage[T1]{fontenc}    
\usepackage{hyperref}       
\usepackage{url}            
\usepackage{booktabs}       
\usepackage{amsmath, amsfonts}       
\usepackage{mathtools}
\usepackage{lipsum}		
\usepackage[pdftex]{graphicx}
\usepackage[numbers]{natbib}
\usepackage{doi}
\newcommand{\bm}[1]{\mbox{\boldmath{$#1 $}}} 
\newtheorem{theorem}{Theorem}

\graphicspath{Figure/}

\title{Sparse estimation in ordinary kriging \\ for functional data}

\author{ \href{https://orcid.org/0000-0002-6286-5072}{\includegraphics[scale=0.06]{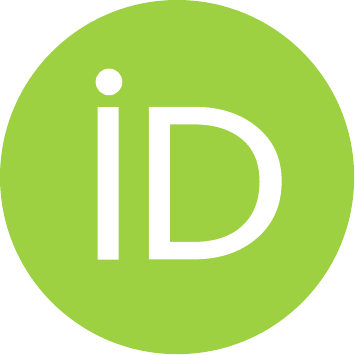}\hspace{1mm}Hidetoshi Matsui} \\
	Faculty of Data Science\\
	Shiga University\\
	1-1-1 Banba, Hikone, Shiga \\
	\texttt{hmatsui@biwako.shiga-u.ac.jp}\\
	\And
	Yuya Yamakawa \\
        Graduate School of Informatics\\
	Kyoto University\\
	Yoshida-Honmachi, Sakyo-ku, Kyoto \\
        \texttt{yuya@kyoto-u.ac.jp}\\
}

\renewcommand{\shorttitle}{Sparse estimation for functional kriging}

\hypersetup{
pdftitle={Sparse estimation for functional kriging},
pdfsubject={q-bio.NC, q-bio.QM},
pdfauthor={Hidetoshi Matsui},
pdfkeywords={First keyword, Second keyword, More},
}

\begin{document}
\maketitle

\begin{abstract}
We introduce a sparse estimation in the ordinary kriging for functional data.  
The functional kriging predicts a feature given as a function at a location where the data are not observed by a linear combination of data observed at other locations.  
To estimate the weights of the linear combination, we apply the lasso-type regularization to minimize the expected squared error.  
We propose an algorithm to derive an estimator using an augmented Lagrange method.  
Tuning parameters included in the estimation procedure are selected by cross-validation.  
Since the proposed method can shrink some of the weights of the linear combination to exactly zero, we can investigate which locations are necessary or unnecessary to predict the feature.  
Simulation and real data analysis show that the proposed method appropriately provides reasonable results.
\end{abstract}

\section{Introduction}

Functional data analysis has received considerable attention in various fields such as medicine, social science, and biology, as one of the techniques that treat longitudinal data rather than scalar data. 
The various methodologies that have been developed include functional regression analysis, functional principal component analysis, and functional canonical correlation analysis \cite{RaSi2005,KoRe2017}.  
More recently, there has been a growing interest in approaches to functional data where individuals are dependent on each other, in particular data not dependent on individuals \cite{KoSt2023}.  
Examples include time series analysis \cite{Bo2000,HyUl2007} and spatial data analysis \cite{DeGiCo_etal2010,Kokoszka2019,MaGi2021} for functional data.  

In the present work, we focus on spatial data analysis for functional data. 
Spatial data analysis considers the problem of analyzing univariate or multivariate data observed at locations distributed over a certain space while considering spatial correlation \cite{GeDiGu_etal2010,cressie2015statistics}.  
Kriging, one of the most typical methods for analyzing spatial data, predicts a feature at a location where data have not yet been observed by a linear combination of the feature at locations where the data have been observed.  
When we have longitudinal data for a feature observed at several locations, spatial functional data analysis treats longitudinal data as functional data and then applies spatial data analysis such as kriging.  
As an earlier work, Goulard and Voltz \cite{GoVo1993} approached this problem by fitting the longitudinal data using a parametric model and then applying kriging to the fitted data, under the restriction that the number of time points be small.  
\cite{GiDeMa2011} proposed the ordinal kriging method for functional data.  
They derived a procedure for estimating the weights of the linear combination under the assumption of weak stationarity.  
With this procedure, how functional data at observed locations relate to the predicted function at the unobserved location can be investigated.
\cite{GiDeMa2010} extended the ordinary functional kriging to the case where the weights vary with time.  
Furthermore, \cite{CaGiMa2013,MeSeDa2013} considered the universal kriging, which loosens the assumption of stationarity.  
Universal kriging assumes that the trend of a functional feature over the space can be modeled by regressing out covariates. 
\cite{AgDuAg2017} also examined predicting a function at an unobserved point using penalized splines with respect to location and time.   

Kriging with sparse regularization enables us to select information relevant to the prediction of the feature at unobserved locations.
Sparse regularization is one of the most useful tools for variable selection in that it estimates the unknown parameters based on regularization with an $L_1$-type penalty \cite{HaTiWa2015}.  
Typical $L_1$-type penalties include the lasso \cite{Ti1996}, elastic net \cite{ZoHa2005} and minimax concave penalty \cite{Zh2010}.  
Sparse regularization has also been applied to kriging methods for scalar data.
\cite{liu2014sparsity} applied a sparsity-inducing penalty to the ordinary kriging, whereas \cite{Hu2011} applied lasso to estimate the coefficients of the linear combination of variables that expresses the trend of the universal kriging.  
\cite{Yang2014}, \cite{ZhYaYe_etal2019}, and \cite{Pa2021} also looked at using lasso for selecting important variables determining the trend of the universal kriging.  

If we can estimate kriging weights as exactly zero, then we can investigate which locations are related to predicting a feature at an unobserved location.  
To obtain such estimators, we propose applying sparse regularization to the weights of the ordinary kriging for functional data.  
We obtain sparse estimators of the weights by minimizing the integrated expected squared error with the $L_1$-type penalty under the assumption of second-order stationarity for the feature.  
Here we apply an adaptive lasso-type regularization \cite{Zo2006} to incorporate the ordinary kriging estimator.  
Since it is difficult to derive estimators of kriging weights with the $L_1$-type penalty, we apply an augmented Lagrange method to obtain them numerically.  
We also show that the algorithm converges to the optimal solution.  
Tuning parameters included in the estimation are selected by cross-validation.  
We apply the proposed method to the analysis of simulated data and weather data and investigate whether the proposed method gives appropriate estimates.  

The remainder of this paper is organized as follows.  
Section 2 provides the ordinary kriging for functional data as background.
In Section 3 we introduce sparse estimation for the ordinary functional kriging.  
Section 4 shows the results of simulation and real data analysis.  
Finally, we conclude with the main points in Section 5.  

\section{Ordinary kriging for functional data}
Let $X(\bm s; t)$ be a random function of $\bm s\in \mathcal S\subset \mathbb R^d$ and $t\in\mathcal T\subset \mathbb R$ square-integrable with respect to $t$.  
Typically, $\bm s$ and $t$ correspond to location and time, respectively.  
Suppose that $X(\bm s; t)$ satisfies the second-order stationarity and isotropy conditions with respect to $\bm s$, that is, 
\begin{gather*}
    {\rm E}\left[X(\bm s; t) \right] = m(t), \\
    {\rm Cov}\left[X(\bm s+\bm h; t), X(\bm s; u)\right] = C(\|\bm h\|; t, u)
\end{gather*}
for all $\bm h\in \mathcal S$, where $m(t)$ is a function that depends only on $t$ and $C(r; t, u)$ is a function that depends only on $r = \|\bm h\|$, $t$, and $u$.
Under the above conditions, the function   
\begin{align*}
    \gamma(r; t) 
    = \frac{1}{2}{\rm V}[X(\bm s+\bm h; t) - X(\bm s; t)]
    = \frac{1}{2}{\rm E}[X(\bm s+\bm h; t) - X(\bm s; t)]^2
\end{align*}
depends only on $r=\|\bm h\|$ and $t$.
In this context, $\gamma(r; t)$ is called a semi-variogram, and its integral $\gamma(r) = \int_{\mathcal T}\gamma(r; t)dt$ is called a trace-variogram. 

We let $X(\bm s_1; t), \ldots, X(\bm s_n, t)$ denote random functions at locations $\bm s_1, \ldots, \bm s_n \in \mathcal S$, respectively, where we assume $\bm s_i \neq \bm s_j$ $(i\neq j)$.  
The aim of the ordinary kriging used here is to predict a function $X(\bm s_0; t)$ at location $\bm s_0 \neq \bm s_i$ $(i=1,\ldots, n)$ based on the data observed at $\bm s_1, \ldots, \bm s_n$.
We assume that the predicted function for $X(\bm s_0; t)$ is expressed as a linear combination of the functions at the observed points:
\begin{align}
\label{eq:xhat}
    \widehat X(\bm s_0; t) = \sum_{i=1}^n \lambda_i X(\bm s_i; t),
\end{align}
where $\lambda_1, \ldots, \lambda_n$ are the unknown kriging weights.  
Equation \eqref{eq:xhat} allows us to quantify how and which functional feature at the observed locations contributes to the prediction at the unobserved location.  
Under the condition that the function $\widehat X(\bm s_0; t)$ satisfies second-order stationarity, $\lambda_1, \ldots, \lambda_n$ satisfy $\sum_{i=1}^n \lambda_i = 1$.

In the ordinary kriging for functional data, we estimate $\lambda_i$ in \eqref{eq:xhat} by solving the following constrained optimization problem \cite{GiDeMa2011}:
\begin{align*}
    \begin{aligned}
        & \mbox{Minimize} & & \int_{\mathcal T} E\left[ X(\bm s_0; t) - \hat X(\bm s_0; t) \right]^2dt
        \\
        & \mbox{subject to} & & \sum_{i=1}^n \lambda_i -1 = 0.
    \end{aligned}
\end{align*}
Solving the above problem is reduced to minimizing the following Lagrange function with respect to $\bm{\lambda}$ and $\mu$ by the method of Lagrange multipliers:
\begin{align}
    L(\bm\lambda, \mu) &= \int_{\mathcal T} E\left[ X(\bm s_0; t) - \hat X(\bm s_0; t) \right]^2dt
    +2\mu\left(\sum_{i=1}^{n}\lambda_i-1 \right) \nonumber\\
    &= \bm\lambda^\top C \bm\lambda - 2\bm c_0^\top \bm\lambda + 2\mu(\bm 1_n^\top \bm\lambda - 1).
    \label{eq:ofk}
\end{align}
where $\mu \in \mathbb{R}$ is a Lagrange multiplier, $\bm\lambda = (\lambda_1, \ldots, \lambda_n)^\top$, $C = (\int_{\mathcal T}C(\|\bm s_i-\bm s_j\|; t, t)dt)_{ij}$, $\bm c_0 = (\int_{\mathcal T}C(\|\bm s_1-\bm s_0\|, t, t)dt, \ldots, \int_{\mathcal T}C(\|\bm s_n-\bm s_0\|; t, t)dt)^\top$, and $\bm 1_n = (1,\ldots, 1)^\top$ is an $n$-dimensional vector.  
Note that $C$ is positive semi-definite, since the matrix whose $(i,j)$-th element is $C(\|\bm s_i-\bm s_j\|; t, t)$ for a fixed $t$ is positive semi-definite.  
We assume more specifically that $C$ is positive definite.  
The solution that minimizes \eqref{eq:ofk} is obtained by simple matrix calculation under the conditions of second-order stationarity and isotropy. 
Since partial derivatives of \eqref{eq:ofk} with respect to $\bm\lambda$ and $\mu$ respectively are
\begin{align*}
    \frac{\partial L(\bm\lambda, \mu)}{\partial \bm\lambda} =  
    2 C\bm\lambda - 2\bm c_0 + 2\mu\bm 1_n, ~~~
    \frac{\partial L(\bm\lambda, \mu)}{\partial \mu} =     
    2\left(\bm 1_n^\top \bm\lambda - 1\right),
\end{align*}
we should solve the following linear system to estimate the ordinary kriging weights $\lambda_{1}, \ldots, \lambda_{n}$:
\begin{align}
	\begin{pmatrix}
		C & \bm 1_n \\
		\bm 1_n^\top & 0
	\end{pmatrix}
	\begin{pmatrix}
		\bm\lambda \\ \mu
	\end{pmatrix}
	= 
	\begin{pmatrix}
		\bm c_0 \\ 
		1
	\end{pmatrix}.
 \label{eq:OFKeq}
\end{align}

However, it is difficult to estimate the function $\int_{\mathcal T}C(r; t, t)dt$ directly, so instead we use the relationship $\gamma(r) = \int_{\mathcal T}C(0; t, t)dt - \int_{\mathcal T}C(r; t, t)dt$ and estimate the trace-variogram $\gamma(r)$.  
Trace-variogram $\gamma(r)$ can be obtained by calculating the sample variance of the difference of pairs of observations such that the distance between two locations equals $r$.  
Since such pairs are very scarce, we calculate the variance using a set of observations such that the distance between two points is close to $r$.  
Denote a set of pairs of points whose distances are close to $r$ as 
\begin{align*}
    N_r = \{(\bm{s}_{i}, \bm{s}_{j}) ; \|\bm{s}_{i} - \bm{s}_{j}\| \approx r\},  
\end{align*}
then estimate the trace-variogram $\gamma(\|\bm s_i - \bm s_j\|)$ by the sample variance using $N_r$:
\begin{align}
    \widehat{\gamma}(\|\bm s_i - \bm s_j\|) = 
    \frac{1}{2|N_r|}\sum_{i,j\in N_r}\int_{\mathcal T} \left[x(\bm s_i; t) - x(\bm s_j; t)\right]^2 dt,
    \label{eq:vario-est}
\end{align}
where $|N_r|$ is the size of $N_r$.
See \cite{cressie2015statistics} for details of the estimation of a variogram.  

In real situations, data are observed longitudinally at discrete time points rather than continuously.
Therefore, we need to transform the observed data into functions.  
To do this, we assume that functional data $x(\bm s_i; t)$ are expressed in terms of a basis expansion as follows.
\begin{align}
    x(\bm s_i; t) = \sum_{m=1}^M w_{im}\phi_m(t) 
    = \bm w_i^\top \bm\phi(t),
    \label{eq:baseexp}
\end{align}
where $\bm\phi(t) = (\phi_1(t), \ldots, \phi_M(t))^\top$ is a vector of basis functions and $\bm w_i = (w_{i1}, \ldots, w_{iM})^\top$ is a vector of coefficients.  
After estimating the coefficient $\bm w$, we treat $\{x(\bm s_i; t) = \bm w_i^\top \bm\phi(t); i=1,\ldots, n\}$ as functional data.  
For details of the method for estimating $\bm w$, see \cite{GrSi1994,RaSi2005}.
Using the basis expansion, the integral in \eqref{eq:vario-est} is calculated as 
\begin{align*}
	\int_{\mathcal T} \left[x(\bm s_i; t) - x(\bm s_j; t)\right]^2 dt = 
	(\bm w_i - \bm w_j)^\top\Phi(\bm w_i - \bm w_j), 
\end{align*}
where $\Phi = \int_{\mathcal T} \bm\phi(t)\bm\phi(t)^\top dt$.  
Then the value of \eqref{eq:vario-est} is obtained by fitting $\widehat\gamma(r)$ by some parametric models such as Gaussian or exponential models, where here we use the Mat\'ern model.  

Therefore, the ordinary functional kriging (OFK) estimators of $\bm\lambda$ and $\mu$ are given by
\begin{align}
	\begin{pmatrix}
		\hat{\bm\lambda} \\ \hat\mu
	\end{pmatrix}
	= 
	\begin{pmatrix}
		C & \bm 1 \\
		\bm 1^\top & 0
	\end{pmatrix}^{-1}
	\begin{pmatrix}
		\bm c_0 \\ 
		1
	\end{pmatrix}.
    \label{eq:solution}
\end{align}
As a result, when we have functional data $x(\bm s_1; t), \ldots, x(\bm s_n; t)$ at $n$ points, we predict a function $x(\bm s_0; t)$ at a point $\bm s_0$ as 
\begin{align}
    \widehat x(\bm s_0; t) = \sum_{i=1}^n \widehat\lambda_i x(\bm s_i; t).
    \label{eq:xhat0}
\end{align}
For more details, see \cite{GiDeMa2011,MaGi2021}.  
\section{Sparse ordinary kriging for functional data}
In this section, we apply sparse regularization to estimate the weights $\lambda_i$ in \eqref{eq:xhat}.
First, for given hyperparameters $\eta$ and $\tau$, we consider the following minimization problem obtained by imposing an $L_1$-type penalization on the minimization problem~\eqref{eq:ofk}.
\begin{align}
\label{eq:sofk}
    \begin{aligned}
        & \mbox{Minimize} & & f(\bm\lambda) \coloneqq \bm\lambda^\top C \bm\lambda - 2\bm c_0^\top \bm\lambda +\eta\sum_{i=1}^{n}\widehat w_i|\lambda_i|
        \\
        & \mbox{subject to} & & g(\bm\lambda) \coloneqq \bm 1_n^\top \bm\lambda - 1 = 0,
    \end{aligned}
\end{align}
where $\eta$ is a regularization parameter and $\widehat w_i>0$ is an adaptive weight. Here we set $\widehat w_i = |\widehat\lambda_i|^{-\tau}$, where $\widehat\lambda_i$ is the ordinary kriging estimator of $\lambda_i$ obtained by the method described in the previous section, and $\tau>0$ is a tuning parameter. Note that the problem~\eqref{eq:sofk} is obtained by adding an adaptive lasso-type penalty \cite{Zo2006} to the OFK.
We call the solution 
of \eqref{eq:sofk} the sparse ordinary functional kriging (SOFK) estimator.  
\par
Due to the additional constraint for $g(\bm\lambda)$ of the optimization problem \eqref{eq:sofk}, it is difficult to directly apply the well-known estimation algorithms for lasso, such as the coordinate descent method \cite{FrHaTi2010a} and the alternating direction method \cite{BoPaCh_etal2011}.     
We use the following augmented Lagrangian method, which is among the most famous algorithms in general and is used in many fields.
\begin{description}
    \item[Step 0:] Choose a positive sequence $\{ \rho_{k} \}$. Select an initial values of kriging weights $\bm\lambda_{0}$ and Lagrange multiplier $\mu_{0}$. Set $k \coloneqq 0$.
    \item[Step 1:] Find a minimizer $\bm\lambda^{\ast}$ of the following problem:
    \begin{align} \label{subproblem}
    \begin{aligned}
         & \mbox{Minimize} & & f(\bm\lambda) + \mu_{k} g(\bm\lambda) + \frac{\rho_{k}}{2} g(\bm\lambda)^{2}.
    \end{aligned}
    \end{align}
    Set $\bm\lambda_{k+1} \coloneqq \bm\lambda^{\ast}$.
    \item[Step 2:] Update the Lagrange multiplier $\mu_{k}$ as follows:
    \begin{align*}
        \mu_{k+1} \coloneqq \mu_{k} + \rho_{k} g(\bm\lambda_{k+1}).
    \end{align*}
    \item[Step 3:] If the termination criterion is satisfied, then stop. Otherwise, set $k \leftarrow k+1$ and go back to Step 1.
\end{description}
In Step 2, we have to find the minimizer of the problem \eqref{subproblem}. 
We note that this problem can be recast in the following form:
\begin{align} \label{equivalent_subproblem}
    \begin{aligned}
        & \mbox{Minimize} & & \frac{1}{2} \Vert A \bm\lambda - b \Vert^{2} + \eta \sum_{i=1}^{n} \widehat{w}_{i} |\lambda_{i}|,
    \end{aligned}
\end{align}
where $A$ and $b$ are defined by 
\begin{align*}
    A &\coloneqq \left( 2 C + \rho_{k} \bm 1_{n} \bm 1_n^{\top} \right)^{\frac{1}{2}},
    \\
    b &\coloneqq \left( 2 C + \rho_{k} \bm 1_{n} \bm 1_n^{\top} \right)^{-\frac{1}{2}} ( 2 \bm c_{0} + \rho_{k} \bm 1_{n} - \mu_{k} \bm 1_{n}),
\end{align*}
respectively. 
Proximal gradient-type algorithms can effectively solve this kind of optimization problem.
We apply the Fast Iterative Shrinkage Thresholding Algorithm (FISTA)~\cite{BeTe2009} for the problem~\eqref{equivalent_subproblem} which is equivalent to the problem~\eqref{subproblem}.
\par In the following, we provide a theorem regarding the convergence to an optimum of the above augmented Lagrangian method. The proof is given in Appendix~\ref{Appendix:proof}.
\begin{theorem} \label{th:convergence}
Let $\{ \bm{\lambda}_{k} \}$ be a sequence generated by the augmented Lagrangian method applied to problem~\eqref{eq:sofk}. Suppose that a positive sequence $\{ \rho_{k} \}$ satisfies that $\sum_{k=0}^{\infty} \rho_{k} = \infty$. Then, any accumulation point of $\{ \bm{\lambda}_{k} \}$ is an optimal solution of~\eqref{eq:sofk}.
\end{theorem}


In the present work, the parameter sequence $\{ \rho_{k} \}$ is updated by the following rule:
\begin{align}
    \rho_{k+1} \coloneqq \left\{
    \begin{aligned}
    & \kappa\rho_{k} & & \mbox{if} ~ g(\bm\lambda_{k+1}) > \alpha g(\bm\lambda_{k}), \\
    & \rho_{k} & & {\rm otherwise},
    \end{aligned}
    \right.
    \label{eq:rho}
\end{align}
where $\alpha\in (0,1)$ and $\kappa$ are additional hyperparameters. This updating rule is general and has been adopted in previous studies, such as \cite{BiMa12,LuWuCh12,YaSa22}. Here we fixed the hyperparameters as $\alpha=0.9$ and $\kappa=2$. Note that the sequence $\{ \rho_{k} \}$ defined by~\eqref{eq:rho} with these parameters satisfies the assumption $\sum_{k=0}^{\infty} \rho_{k} = \infty$ in Theorem~\ref{th:convergence}.

In the proposed sparse regularization method, we obtain several minimizers $\bm\lambda^{(1)}, \ldots, \bm\lambda^{(N)}$ for given hyperparameters $(\eta, \tau) = (\eta^{(1)}, \tau^{(1)}), \ldots, (\eta^{(N)}, \tau^{(N)})$, respectively. The best tuning parameters $(\eta, \tau)$ are selected using cross-validation.
\begin{align}
    {\rm CV} = \sum_{i=1}^n\int \left\{x(\bm s_i; t) - \widehat x^{(-i)}(\bm s_i; t)\right\}^2dt,  
    \label{eq:CV}
\end{align}
where $\widehat x^{(-i)}(\bm s_i; t)$ is a function predicted by SOFK at the $i$-th location $\bm s_i$ obtained using the functional data of size $n-1$ excluding the $i$-th observation out of $n$. 
In particular, we adopt $(\eta, \tau)$ to minimize \eqref{eq:CV} as the best tuning parameters.  

The algorithm for estimating $\bm\lambda$ and $\mu$ is summarized as follows.  
\begin{description}
\item[Step 0:] Give a fixed hyperparameter set $\{ (\eta^{(j)}, \tau^{(j)}) \}_{j=1}^{N}$. Set $j \coloneqq 1$.
\item[Step 1:] Find the optimal solution $\bm\lambda^{(j)}$ of \eqref{eq:sofk} with $(\eta, \tau) \coloneqq (\eta^{(j)}, \tau^{(j)})$.
\item[Step 2:] Calculate \eqref{eq:CV} using the estimated $\bm\lambda^{(j)}$ and obtain ${\rm CV}^{(j)}$ using \eqref{eq:CV}.
\item[Step 3:] If $j<N$, then $j \leftarrow j + 1$ and go back to Step~1. Otherwise, select $(\eta^{(\ell)}, \tau^{(\ell)})$, where $\ell$ satisfies ${\rm CV}^{(\ell)} = \displaystyle\min_{1 \leq j \leq N} {\rm CV}^{(j)}$. Obtain the corresponding estimator of $\bm\lambda^{(\ell)}$.  
\end{description}
\section{Example}
\subsection{Simulation}
We examined the effectiveness of the proposed method through an analysis of simulation data.   
This simulation expresses predictions of functions as \eqref{eq:xhat0} at locations where data have not been observed, and investigates the behavior of the estimated kriging weights $\{\hat\lambda_i; i=1,\ldots, n\}$ in \eqref{eq:solution} and the prediction accuracy.  

\begin{figure}[t]
	\begin{center}
		\includegraphics[width=0.37\hsize]{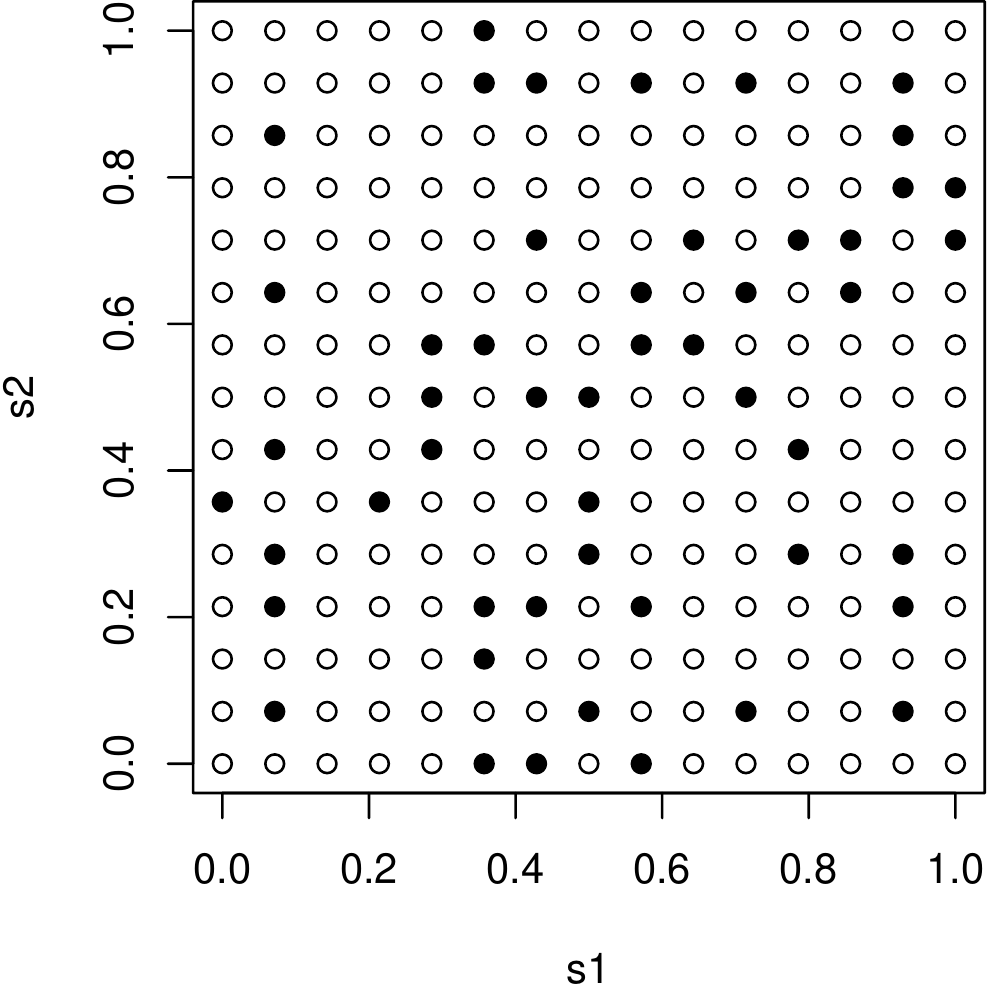}
		\includegraphics[width=0.55\hsize]{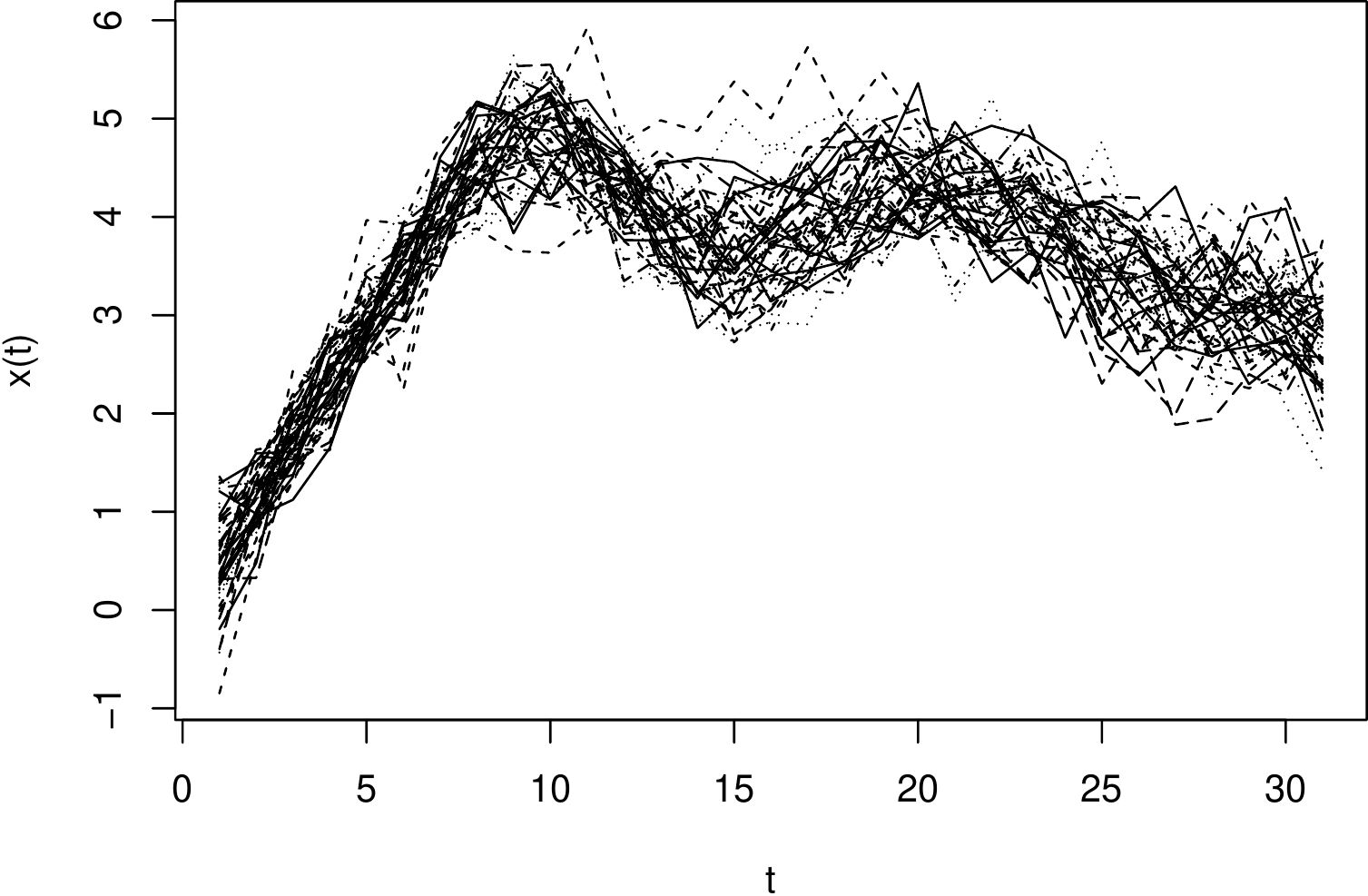}
	\end{center}
	\caption{Examples of simulation data for $n=50$ and ${\rm range}=5$.
 Left: Locations to be investigated. Longitudinal data are observed at black points and not at white points.  
    Right: Observed longitudinal data at black points shown in the left figure.} 
	\label{fig:sim_data}
\end{figure}

The simulation considers a two-dimensional square-shaped domain $\mathcal S = [0,1]\times [0,1]$ and $15\times 15=225$ equally spaced locations in $\mathcal S$.  
Each location has a spatially correlated feature, which is longitudinally observed at $n=15, 25,$ and 50 locations randomly chosen from the $225$ locations, as shown in Figure \ref{fig:sim_data} left.  
We generated longitudinal data $\{x_{ij}; j=1, \ldots, N\}$ at the $i$-th of $n$ locations by adding Gaussian noises to the true structure expressed by basis expansion as
\begin{align*}
    x_{ij} = \bm w_i^\top \bm\phi(t_{ij}) + \varepsilon_{ij}, 
\end{align*}
where $\varepsilon_{ij}$ are independently and identically normally distributed with mean 0 and standard deviation $0.3$. 
Here we used a $B$-spline for $\bm\phi(t)$ with $M=10$ basis functions.  
Coefficients $\bm w_i$ $(i=1, \ldots, n)$ are numerically generated via the variogram model using R package {\tt gstat} \cite{pebesma2004multivariable}, so that the coefficients have positive spatial correlations.  
In generating $\bm w_i$, we assumed an exponential variogram model with the three ranges $1, 5$, and $10$, sill as $2$, and nugget as $0$.  
The number of time points for the longitudinal data is set as $N=31$ for all locations.  
Figure \ref{fig:sim_data} right shows an example of a simulated dataset at the $n$ sites illustrated by the black dots in Figure \ref{fig:sim_data} left.    
We analyzed this dataset by applying the SOFK to predict the features at the $225-n$ locations, where the longitudinal data are not observed, and to investigate the estimated kriging weights.  
We also applied the OFK for comparison.  

As a first process of the analysis, we transformed the observed longitudinal data at $n$ locations by $B$-spline basis functions.  
We used the R package {\tt fda} and fixed the number of basis functions at 10 to restrain the computational cost.
Then we applied SOFK and OFK to each of the test locations, where we used R package {\tt geofd} to calculate the trace-variogram.  
Tuning parameters $\eta$ and $\tau$ were selected by CV in \eqref{eq:CV}.  
We investigated the estimated parameters $\lambda_i$ $(i=1, \ldots, n)$, especially how many parameters are estimated to be zero.  

\begin{table}
    \centering
    \begin{tabular}{ccccc}
    \hline
         &  & \multicolumn{2}{c}{MSE} &  \\
         & range & SOFK & OFK & non-zero\\
    \hline
         & 1     & 5.054 (3.109) & 5.061 (3.106) & 9.160 (3.774) \\
 $n=25$  & 5     & 1.382 (0.790) & 1.397 (0.797) & 6.695 (2.775) \\
         & 10    & 0.852 (0.442) & 0.877 (0.443) & 4.635 (1.717) \\  
    \hline
         & 1     & 3.666 (2.341) & 3.674 (2.326) & 9.989 (4.722)\\
 $n=50$  & 5     & 0.898 (0.473) & 0.899 (0.469) & 7.006 (3.269)\\ 
         & 10    & 0.661 (0.289) & 0.678 (0.295) & 5.669 (2.545)\\  
    \hline
         & 1     & 2.712 (1.461) & 2.727 (1.485) & 12.808 (4.790)\\
 $n=100$ & 5     & 0.777 (0.417) & 0.792 (0.428) & 9.760 (4.467)\\  
         & 10    & 0.569 (0.408) & 0.577 (0.399) & 5.288 (1.512)\\  
    \hline
    \end{tabular}
    \caption{Simulation results. }
    \label{tab:sim}
\end{table}
\begin{figure}[t]
	\begin{center}
		\includegraphics[width=0.45\hsize]{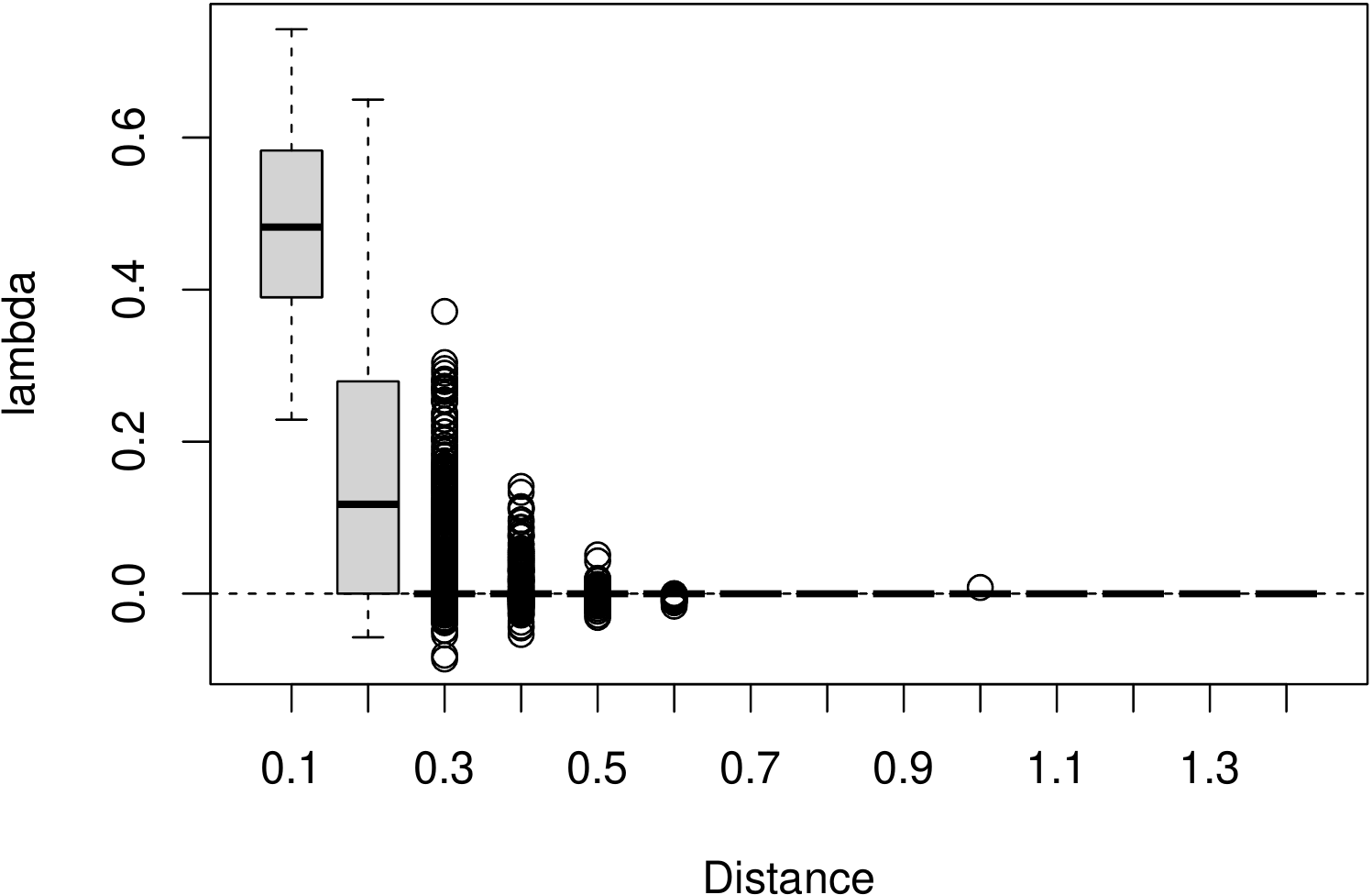} 
		\includegraphics[width=0.45\hsize]{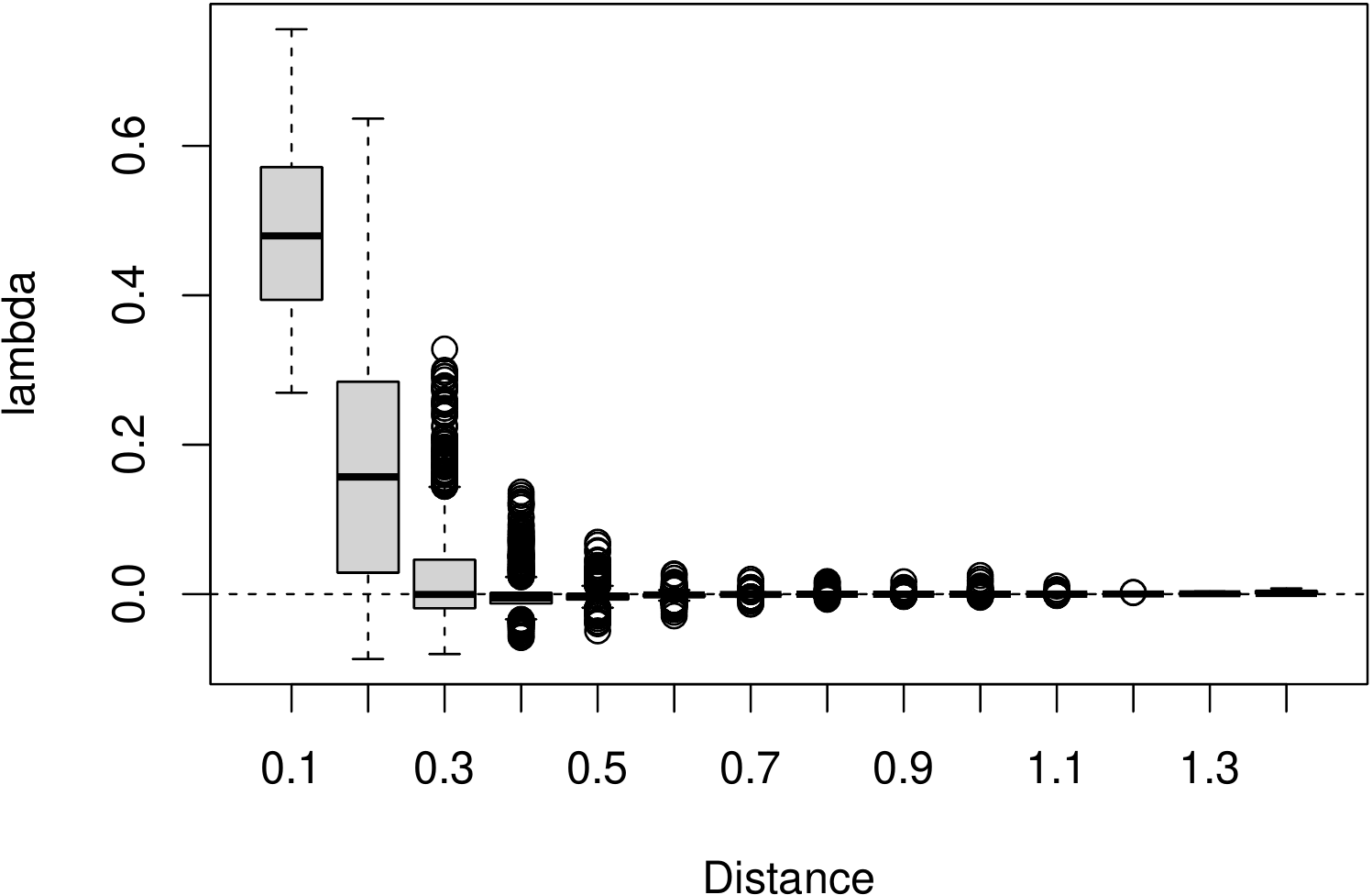}
	\end{center}
	\caption{Box-plots of the SOFK estimators $\hat\lambda_i$ for all unobserved locations according to the distance from the prediction point for $n=50$ and range=10. Left: SOFK, Right: OFK.}
	\label{fig:sim_boxplot} 
\end{figure}
\begin{figure}[t]
	\begin{center}
		\includegraphics[width=0.32\hsize]{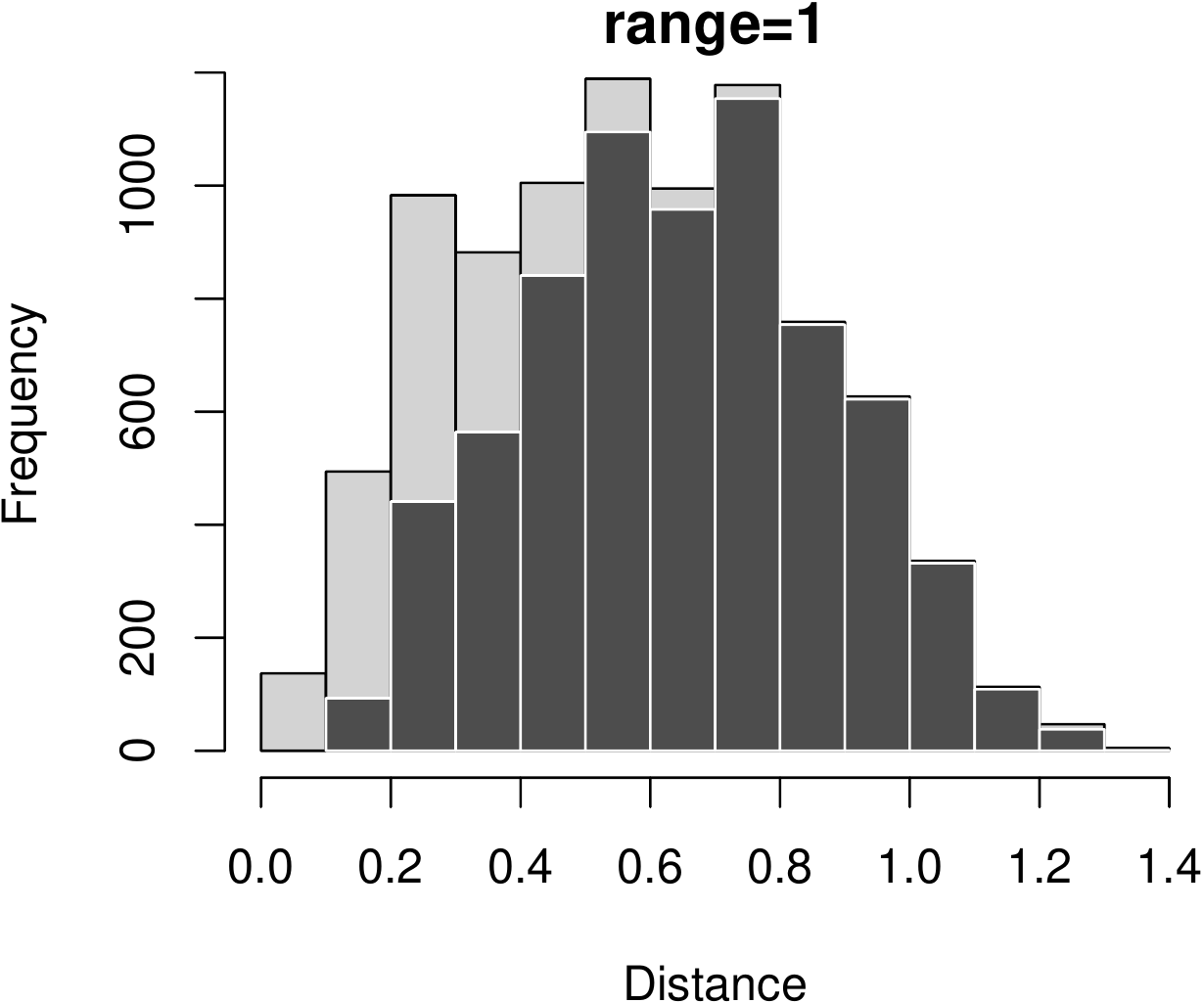}
		\includegraphics[width=0.32\hsize]{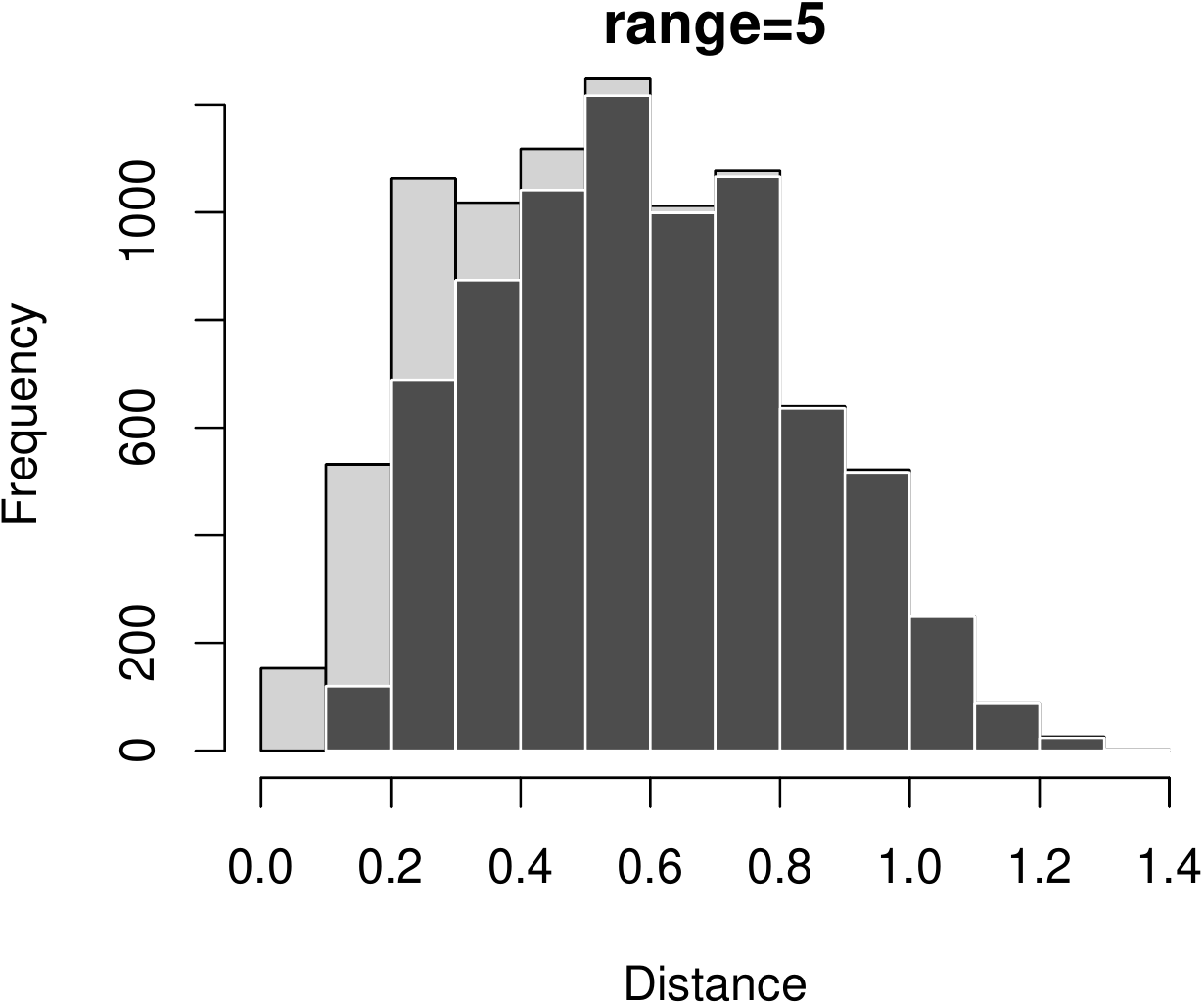}
		\includegraphics[width=0.32\hsize]{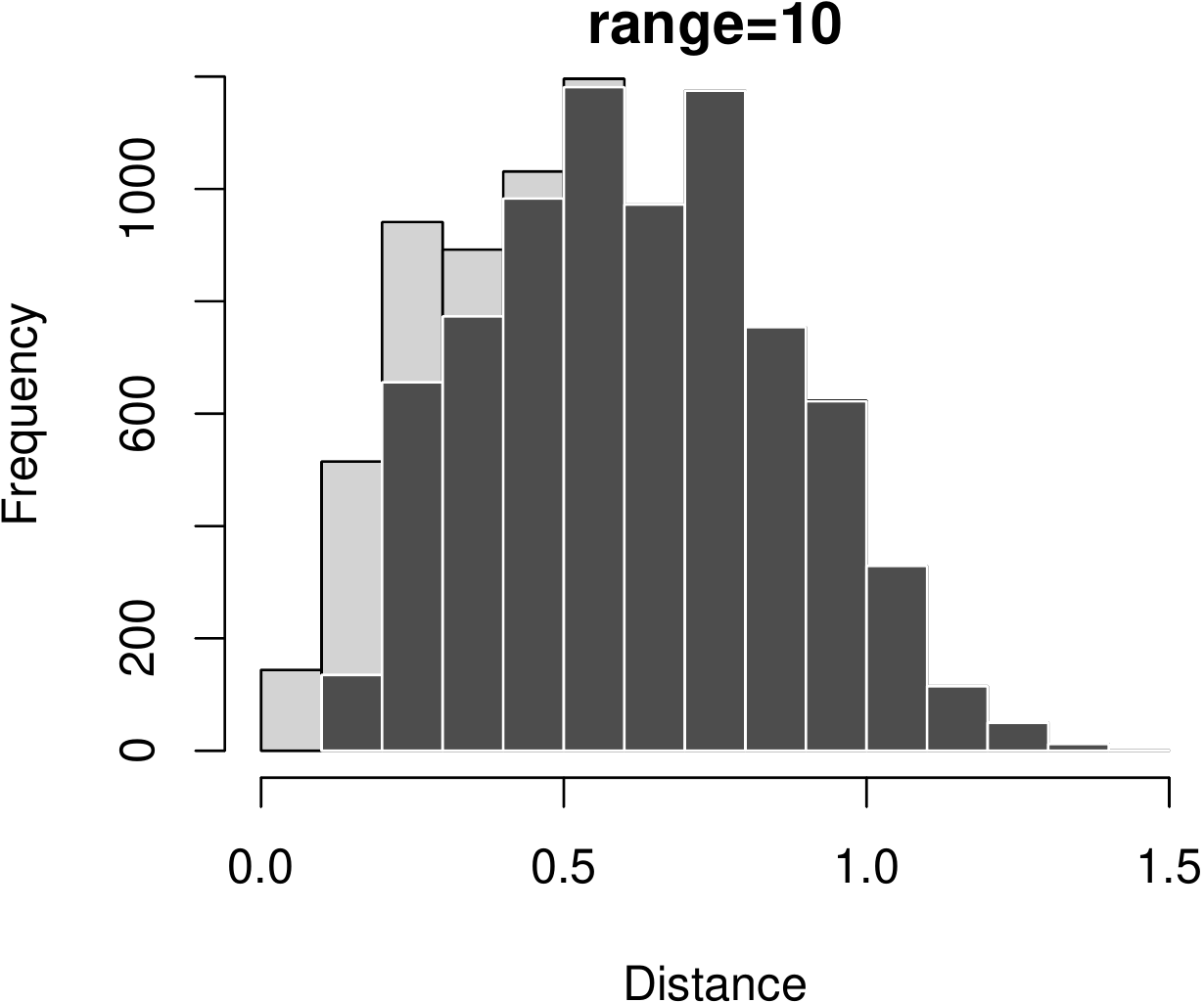} 
	\end{center}
	\caption{Histogram for the frequency of parameters estimated to be non-zero (light gray) and zero (charcoal gray) for $n=50$. }
	\label{fig:sim_hist} 
\end{figure}

Table \ref{tab:sim} shows MSEs and their standard deviations for SOFK and OFK and averages and standard deviations of the numbers of non-zero estimates of the parameters $\lambda_i$ obtained by SOFK.  
This table indicates that for almost all cases the proposed SOFK gives smaller or competitive MSEs than OFK. 
The standard deviations of MSEs become smaller as the sample size grows, except for the case $n=100$ and range $=10$.  
Larger ranges also tend to give stable MSEs, which seems to be due to the larger numbers of zero estimates of $\lambda_i$. 
SOFK shrinks some of the parameters to exactly zero, especially for the case with the larger range.  
This result indicates that data with a larger spatial correlation can be estimated with fewer parameters.  

Figure \ref{fig:sim_boxplot} shows boxplots for estimated parameters $\hat\lambda$ obtained by SOFK and OFK for $n=50$ and ${\rm range}=10$ according to the distance from the locations to be predicted.  
These figures show that the SOFK shrank the parameters to exactly zero at some distance from the predictor point.  
In addition, Figure \ref{fig:sim_hist} shows histograms of numbers of locations and those estimated to be exactly zero according to the distance from the locations to be predicted for all three ranges for $n=50$. 
The weights are never shrunken to exactly zero at the nearest locations, whereas almost all weights are estimated to be zero at locations a short distance away from the test locations.
In particular, we can predict the function at unobserved locations using the functions only for the nearer points as the range increases.  



\subsection{Real data analysis}
\begin{figure}[t]
	\begin{center}
		\includegraphics[width=0.45\hsize]{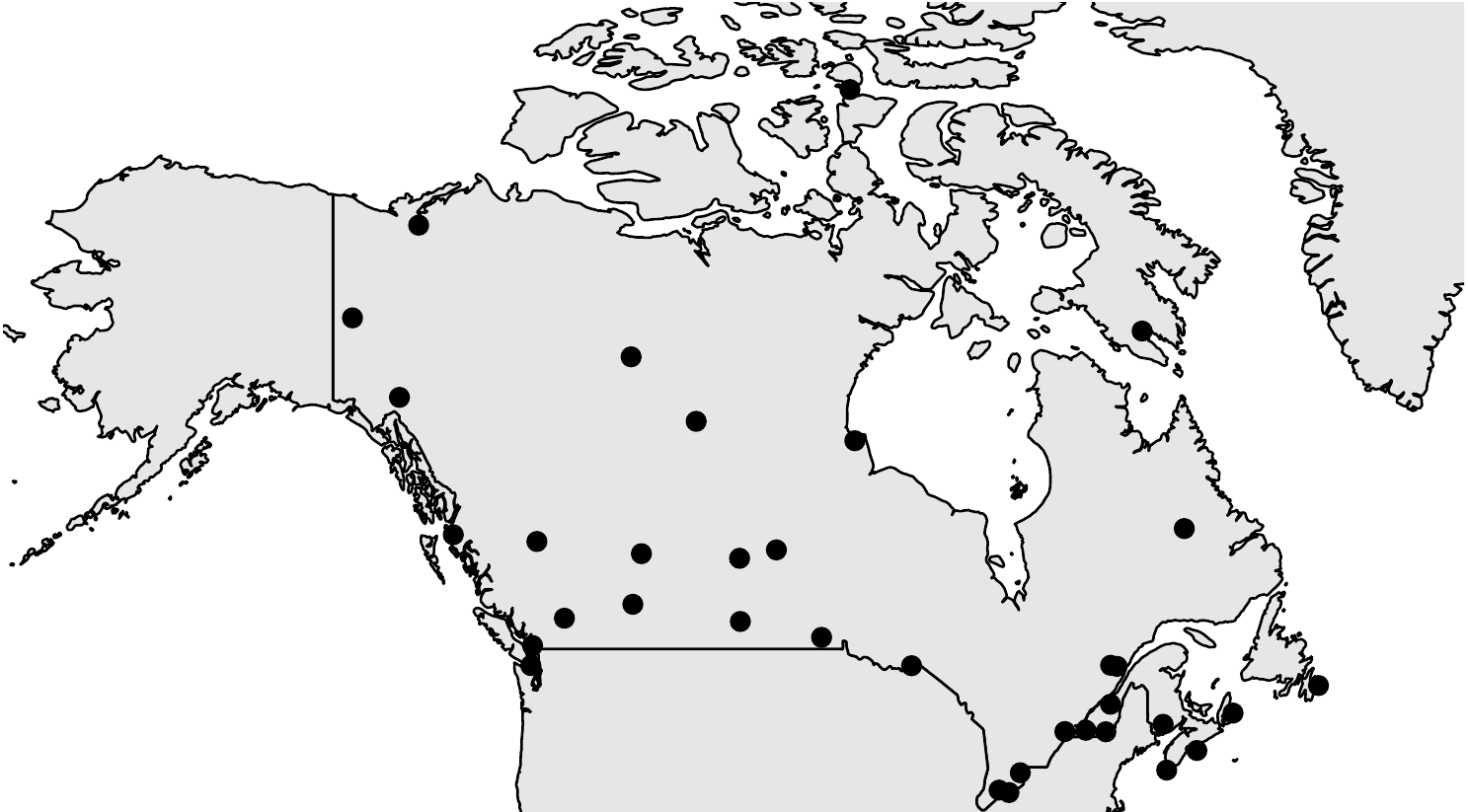}
		\includegraphics[width=0.45\hsize]{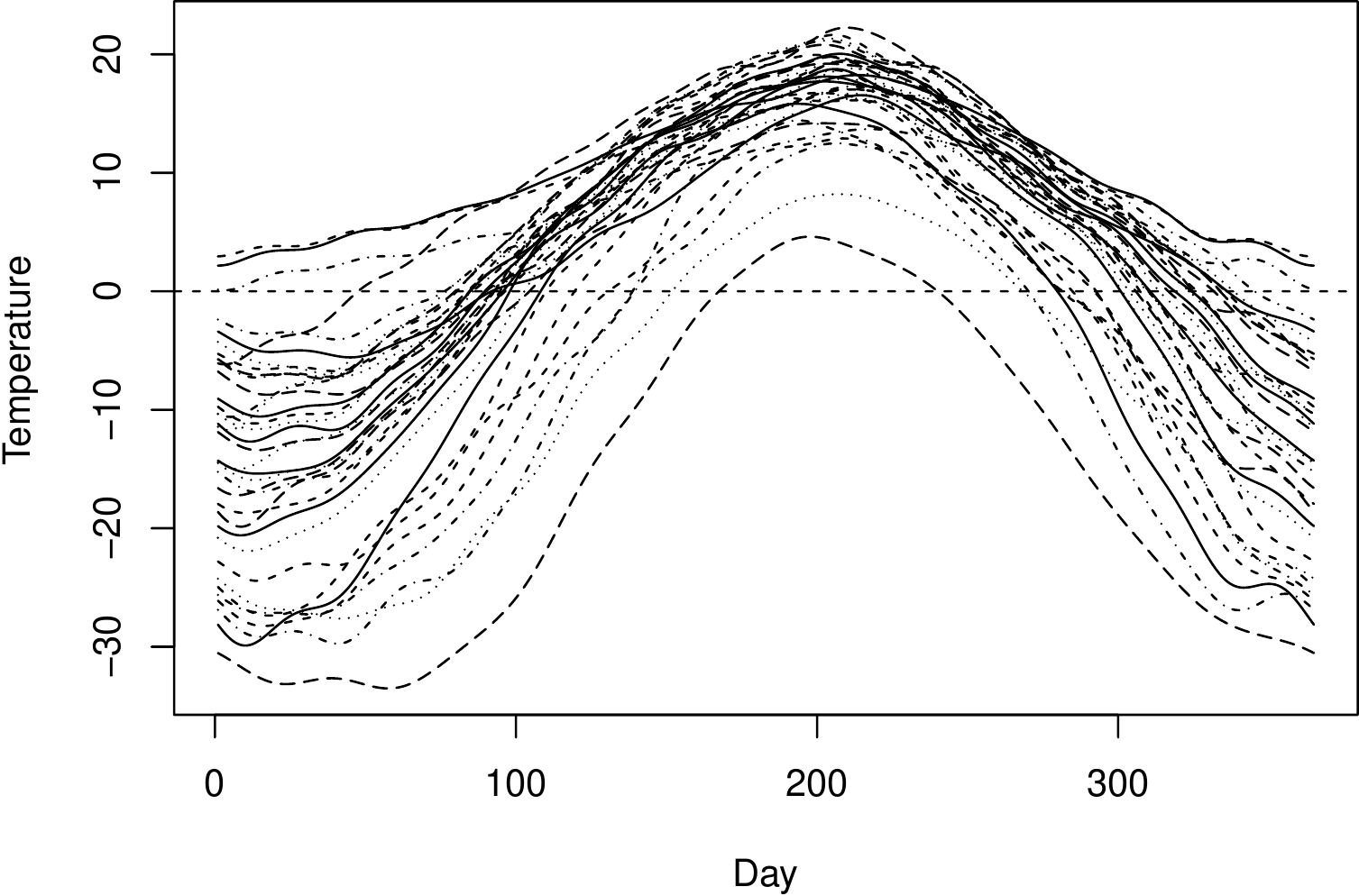}
	\end{center}
	\caption{Canadian weather data.  Left: Observed locations (black circle).  Right: Yearly temperatures at observed 35 locations.} 
	\label{fig:temp_loc}
\end{figure}

We applied the proposed method to the analysis of Canadian weather data, available in R package {\tt fda}.  
The dataset contains locations (longitude and latitude) of 35 cities in Canada and their daily temperatures each of which is averaged from 1960 to 1994.  
The locations and temperature data are shown in Figure \ref{fig:temp_loc}.  
This analysis predicts the annual temperature curve at one location using the annual temperature curves at the remaining 34 locations and investigates which location temperatures are important for this prediction. 
For this analysis, we applied the proposed SOFK.  

First, we transformed the observed temperature data at 35 locations including the location to be predicted into functions by basis expansions.  
Here we used the Fourier series for the basis functions since the data have periodicity, and fixed the number of basis functions 25.  
For this dataset, we calculated the empirical trace-variogram and then estimated the weights $\lambda_i$ in \eqref{eq:xhat} by SOFK.  
Tuning parameters $\eta$ and $\tau$ included in \eqref{eq:sofk} are selected by CV.  
We repeated the analysis, alternating the role of the forecast points.

\begin{figure}[t]
	\begin{center}
		\includegraphics[width=0.45\hsize]{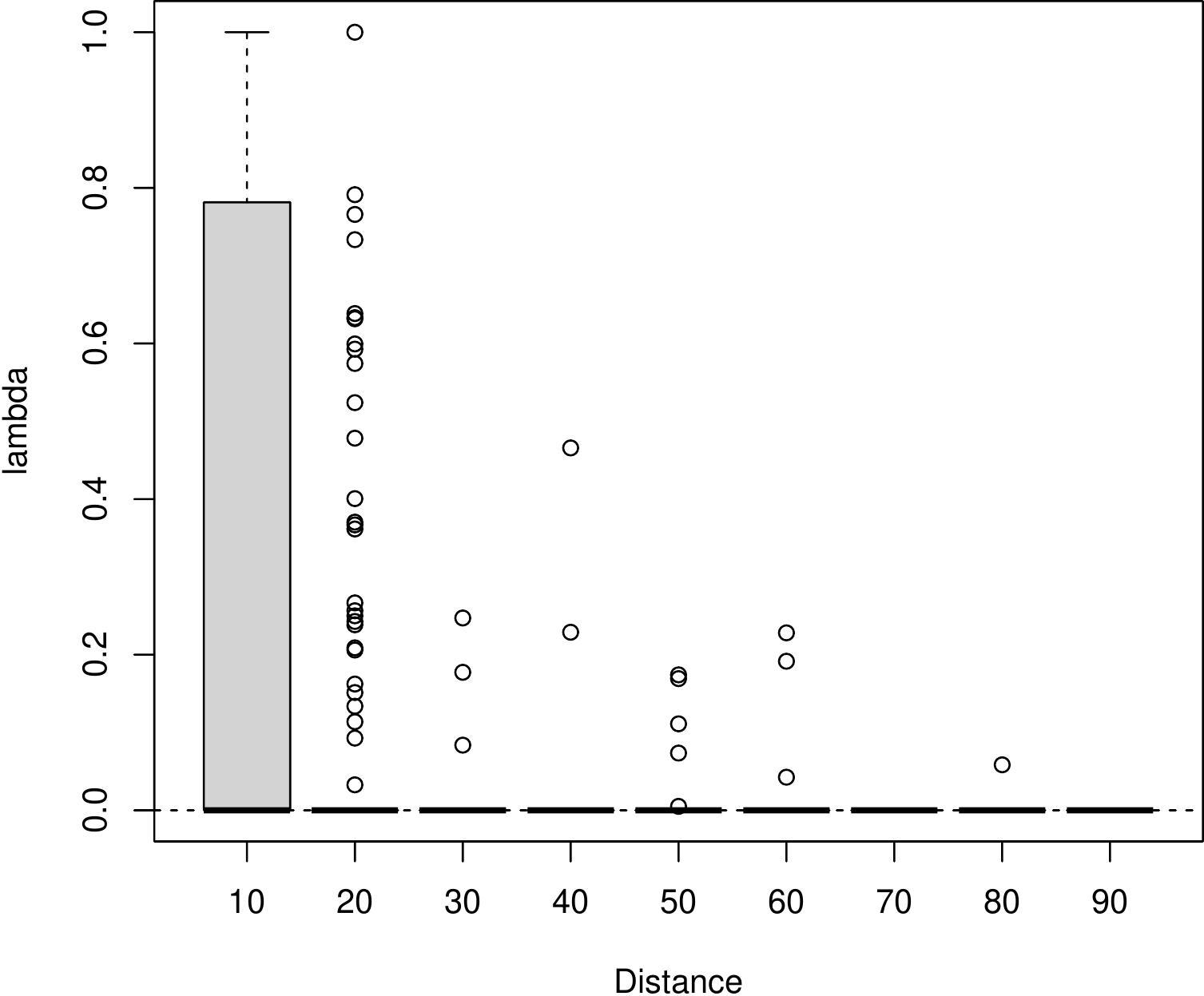}
		\includegraphics[width=0.45\hsize]{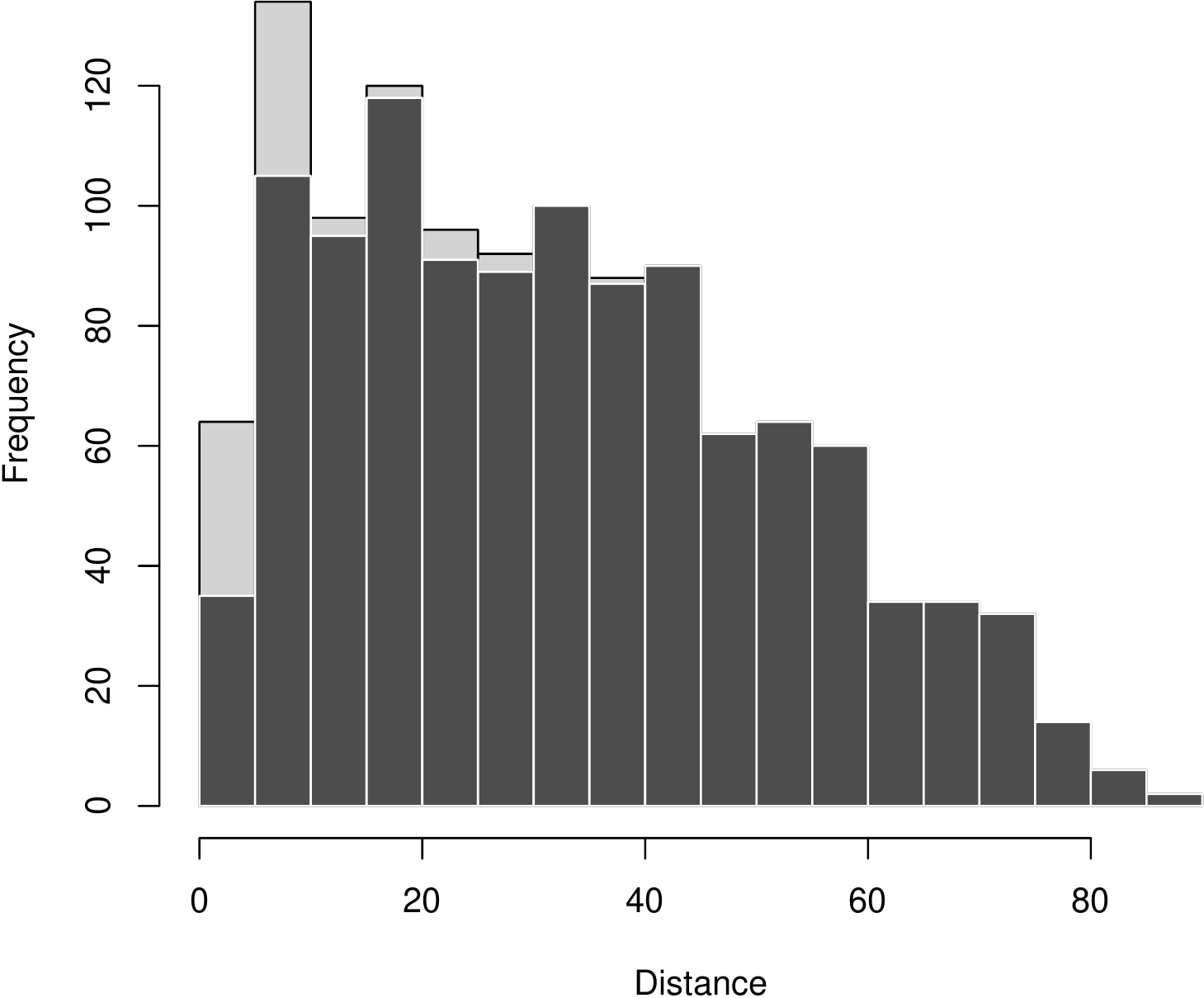}
	\end{center}
	\caption{Left: Boxplots of the estimated SOFK weights $\lambda_1, \lambda_n$ according to the distance from the location to be predicted (horizontal axis).  Right: Histogram of the locations used for the prediction according to the distance from the location to be predicted (horizontal axis).  Black bars are numbers of locations where the kriging weights $\lambda_i$ are estimated to be 0.  
 } 
	\label{fig:temp_boxhist}
\end{figure}

\begin{figure}[t]
	\begin{center}
		\includegraphics[width=0.95\hsize]{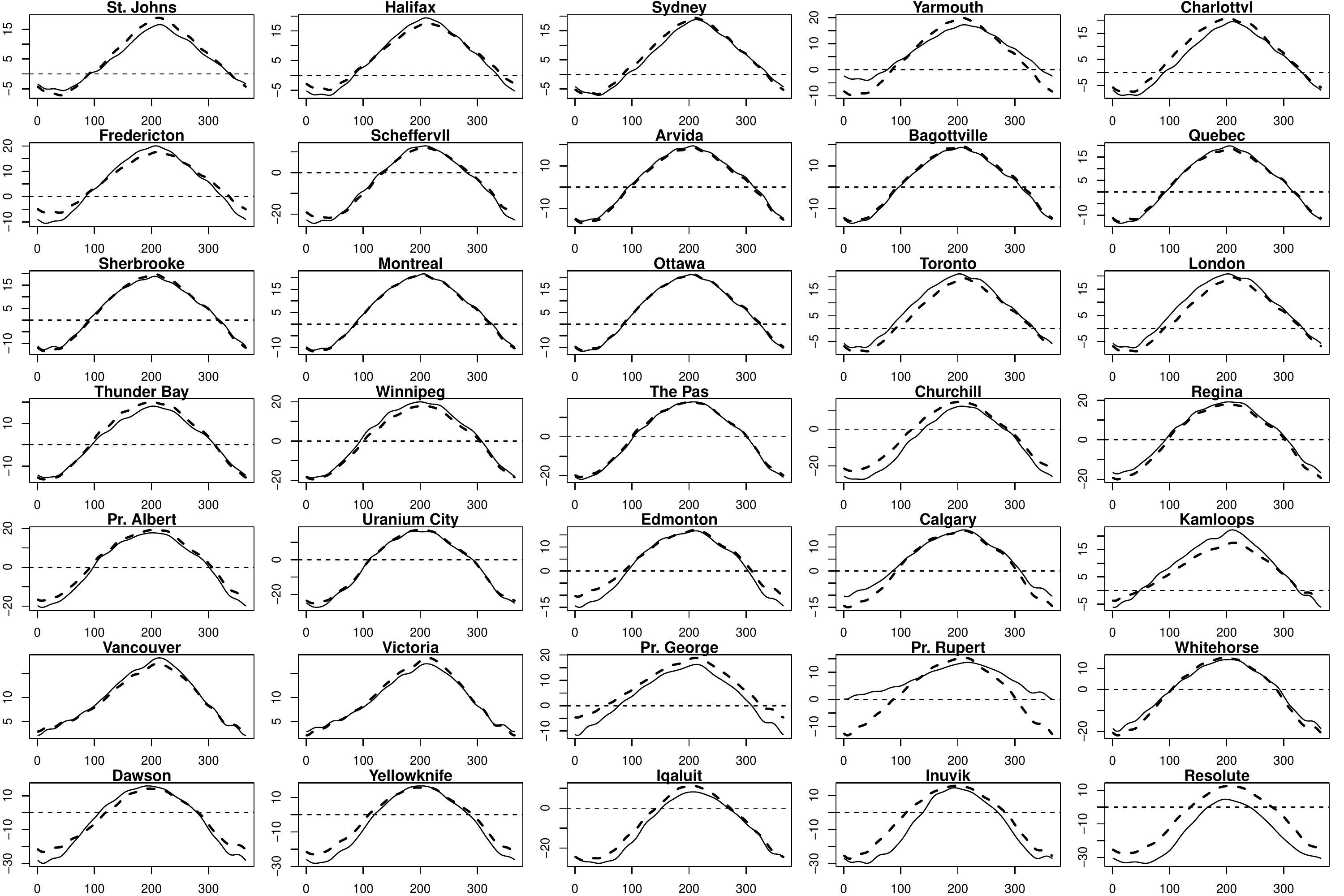}
	\end{center}
	\caption{Prediction results.  Solid: transformed into functional data.  Dashed: estimated by SOFK.} 
	\label{fig:temp_pred}
\end{figure}

Figure \ref{fig:temp_boxhist} left is a box-plot of the SOFK estimators $\hat\lambda_i$ obtained for the prediction of all locations according to the distance from the locations to estimate, and Figure \ref{fig:temp_boxhist} right is the distribution of parameters estimated as zeros or non-zero.  
We can see from these figures that to predict the temperature curve in any city, we only need to use the temperature curves of the locations near that city.  
Figure \ref{fig:temp_pred} shows predicted temperature curves for all cities.  
Temperatures in most cities are appropriately predicted, whereas the remaining ones are poorly predicted.  
The latter may not be suitable for having their curves predicted using a small number of other cities because each of these cities is isolated from the cities where data are observed (e.g., Pr. Rupert and Resolute).  

\begin{figure}[t]
	\begin{center}
		\includegraphics[width=0.47\hsize]{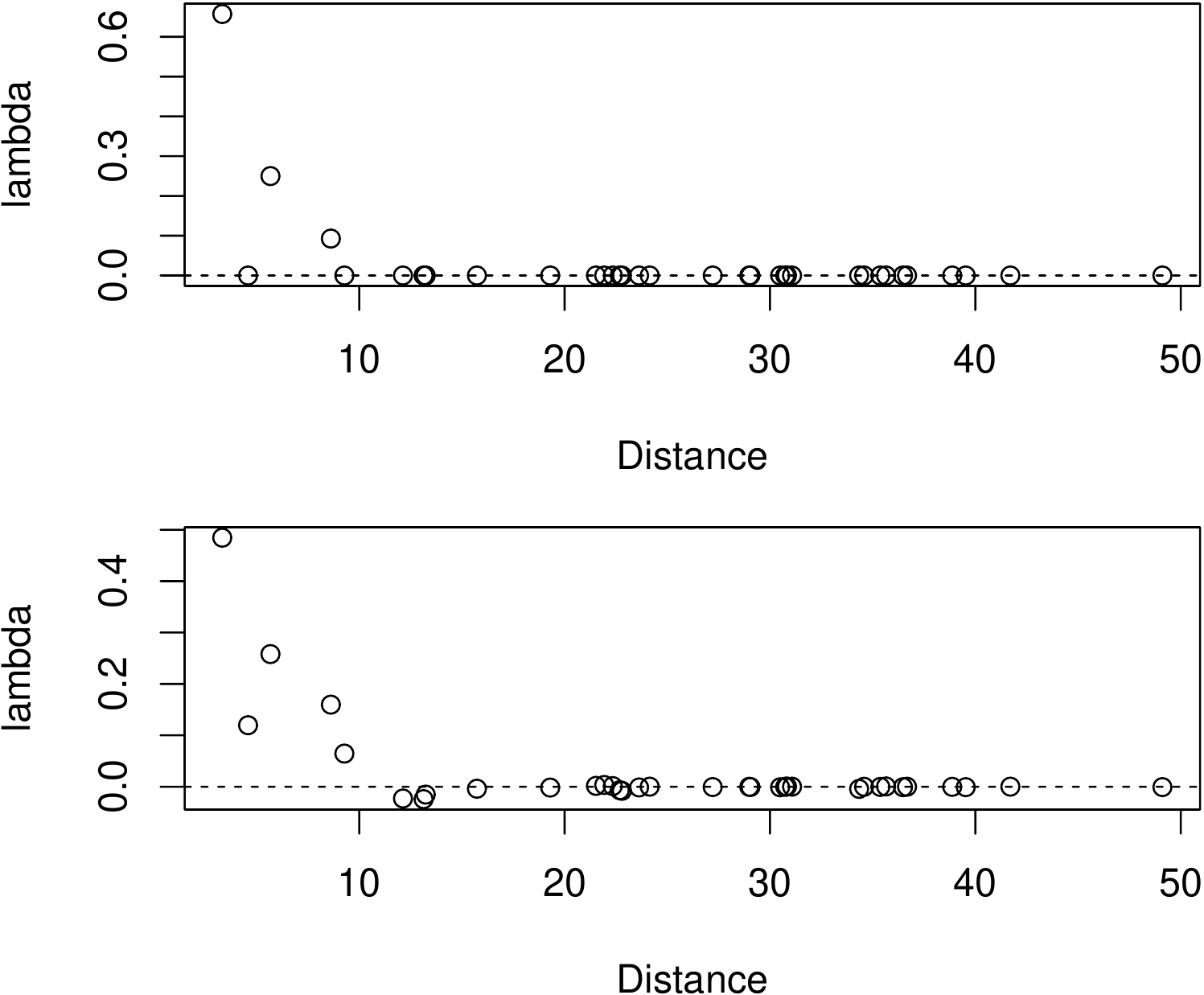}
		\includegraphics[width=0.47\hsize]{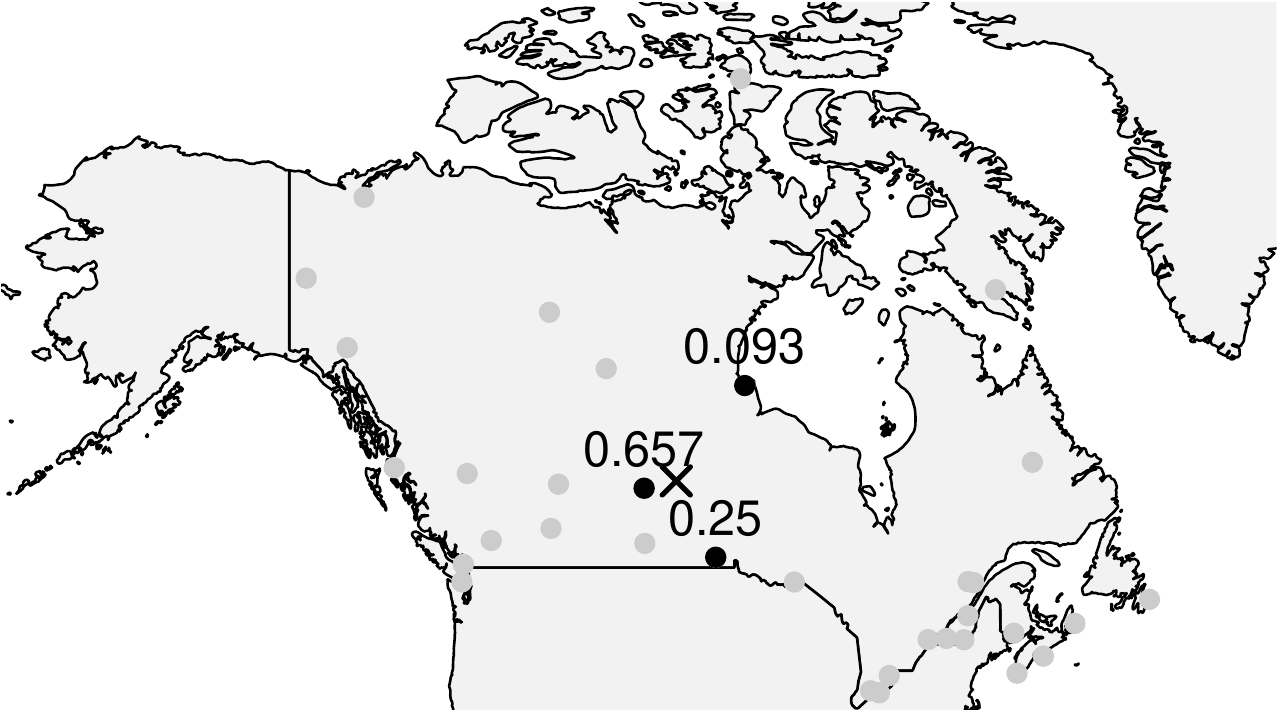}
	\end{center}
	\caption{Estimated kriging weights for predicting the weather at The Pas.  
    Left: Estimated $\lambda_i$ by SOFK (top) and OFK (bottom). The horizontal axis is the distance from The Pas.  Right: Non-zero values of $\lambda_i$ estimated by SOFK in the map of Canada. The location of The Pas is marked as $\times$.} 
	\label{fig:temp_est}
\end{figure}

Parameters $\lambda_1, \ldots, \lambda_n$ estimated by SOFK and OFK for predicting the temperature curve at The Pas are shown in Figure \ref{fig:temp_est} as an example of the prediction.  
We can see that SOFK shrinks the weights of most locations except for three locations near The Pas to exactly zero, whereas OFK does not--- Prince Albert $(0.657)$, Winnipeg $(0.250)$, and Churchill $(0.093)$--- as depicted in Figure \ref{fig:temp_est} right.  
Specifically, the nonzero estimates of SOFK are all positive.  
Although some of the estimates of OFK are negative due to the constraint $\sum_{i=1}^n \lambda_i=1$, this result is against the intuition that temperatures of close locations are similar.  

%
\section{Conclusion}
We have proposed a sparse estimation method for ordinary functional kriging. 
The kriging gives predictions of the feature at unobserved locations by a linear combination of the data at observed locations.  
Using sparse regularization to estimate the weights, we can investigate which combination of locations is used to predict the feature of unobserved locations.  
This result may be applicable to the decision-making of prediction.  
Sparse ordinary functional kriging estimates are obtained by adaptive lasso-type regularization and the augmented Lagrange method.  
Simulation and real data analysis show that the proposed method appropriately estimates the unknown parameters and selected locations that are necessary to predict the feature at an unobserved location.  

The proposed method assumed second-order stationarity, but this constraint is too strong to predict temperatures in all areas.
For example, temperatures in Japan are observed over a wide range from northeast to southwest and it is unnatural to assume stationarity.
To overcome this problem, an extension to the universal kriging that loosens the stationarity assumption should be considered.  
Other sparsity-inducing penalties such as SCAD and MCP may be suitable for performing SOFK.  
A comparison of these penalties remains as future work.  
\bibliographystyle{plain}

\appendix
\def\thesection{\Alph{section}} 
\section{Proof of Theorem~\ref{th:convergence}} \label{Appendix:proof}
In this appendix, we provide a proof of Theorem~\ref{th:convergence}. In the following, we denote the closer and relative interior of a set $S$ as ${\rm cl}(S)$ and ${\rm ri}(S)$, respectively. Moreover, we say that a function $F \colon \mathbb{R} \to \mathbb{R}$ is proper if ${\rm dom}(F) \not = \emptyset$ and $f(x) > -\infty$ for all $x \in \mathbb{R}$, where ${\rm dom}(F) \coloneqq \{ x \in \mathbb{R}; F(x) < \infty \}$.
\\
\\
\noindent {\it Proof.}
We begin by defining some notation:
\begin{align*}
    p(u) \coloneqq \inf \{ f(\bm{\lambda}); \bm\lambda \in {\cal S}(u) \}, \quad {\cal S}(u) \coloneqq \{ \bm\lambda \in \mathbb{R}^{n}; \bm{1}_{n}^{\top}\bm\lambda - 1 = u \}.
\end{align*}
From~\cite[Proposition~5.2.1]{Bertsekas2015}, it is sufficient to show that 
\begin{itemize}
\item[(i)] $p(0)$ is finite;
\item[(ii)] the function $p$ is closed and proper;
\item[(iii)] the dual problem of \eqref{eq:sofk} has at least one optimal solution.
\end{itemize}
Note that $f$ is bounded below on $\mathbb{R}^{n}$ because 
\begin{align} 
- \bm{c}_{0}^{\top} C^{-1} \bm{c}_{0} 
&= 
\Vert C^{\frac{1}{2}} \bm\lambda - C^{-\frac{1}{2}} \bm{c}_{0} \Vert^{2} - \bm{c}_{0}^{\top} C^{-1} \bm{c}_{0} \nonumber
\\
&\leq \Vert C^{\frac{1}{2}} \bm\lambda - C^{-\frac{1}{2}} \bm{c}_{0} \Vert^{2} - \bm{c}_{0}^{\top} C^{-1} \bm{c}_{0} + \eta \sum_{i=1}^{n} \widehat{w}_{i} | \lambda_{i} | \nonumber
\\
&= f(\bm\lambda)  \qquad \forall \bm\lambda \in \mathbb{R}^{n}. \tag{A.1}
\end{align}
Note also that $\frac{1}{n}\bm{1}_{n} \in {\cal S}(0)$. Hence, we have from the definition of $p(0)$ that $p(0) \leq f(\frac{1}{n}\bm{1}_{n})$. Thus, it is clear that $p(0)$ is finite, that is, that item~(i) is verified.
\par
From here, we show item~(ii). Since $p$ is proper from item~(i) and inequality~(A.1), we will prove that $p$ is closed. 
We first show that ${\rm dom}(p)$ is closed. Let us take $v \in \mathbb{R}$ arbitrarily. 
We note that $\bm{\lambda}_{v} \coloneqq \frac{1+v}{n} \bm{1}_{n} \in {\cal S}(v)$. Thus, we readily have $p(v) \leq f(\bm{\lambda}_{v}) < \infty$, that is, $v \in {\rm dom}(p)$. Since ${\rm dom}(p) = \mathbb{R}$ holds, it is clear that ${\rm dom}(p)$ is closed. Secondly, we prove that ${\cal V}(\gamma) \coloneqq \{ u \in \mathbb{R}; p(u) \leq \gamma \}$ is closed for each $\gamma \in \mathbb{R}$. Let $\gamma \in \mathbb{R}$ and $u \in {\rm cl}({\cal V}(\gamma))$ be arbitrary. By the definition of ${\rm cl}({\cal V}(\gamma))$, there exists $\{ u_{j} \} \subset \mathbb{R}$ such that 
\begin{align} \tag{A.2}
    p(u_{j}) \leq \gamma \quad \forall j \in \mathbb{N}, \qquad \lim_{j\to\infty} u_{j} = u.
\end{align}
We arbitrarily take $j \in \mathbb{N}$. 
From the definitions of $p(u_{j})$ and ${\cal S}(u_{j})$, there exists $\bm\xi_{j} \in \mathbb{R}^{n}$ such that
\begin{align} \tag{A.3}
\bm{1}_{n}^{\top} \bm\xi_{j} - 1 = u_{j}, \qquad f(\bm\xi_{j}) < p(u_{j}) + \frac{1}{j} \quad \forall j \in \mathbb{N}.
\end{align}
Combining~(A.1)--(A.3) gives that $- \bm{c}_{0}^{\top} C^{-1} \bm{c}_{0} \leq f(\bm\xi_{j}) \leq \gamma + 1$, namely that $\{ f(\bm\xi_{j}) \}$ is bounded. It then follows from equality~(A.1) that $\{ \bm\xi_{j} \}$ is also bounded. 
Thus, we can assume without loss of generality that $\bm\xi_{j} \to \bm\xi^{\ast}$ as $j \to \infty$. 
Using $u_{j} \to u ~ (j \to \infty)$ in~(A.2) and $\bm{1}_{n}^{\top} \bm\xi_{j} - 1 = u_{j}$ in~(A.3), we get $\bm{1}_{n}^{\top} \bm\xi^{\ast} - 1 = u$, i.e., $\bm\xi^{\ast} \in {\cal S}(u)$. Then, noting the definition of $p(u)$ yields $p(u) \leq f(\bm\xi^{\ast})$.
Since $f(\bm\xi^{\ast}) \leq \gamma$ holds by exploiting (A.2) and (A.3), we can derive $p(u) \leq \gamma$. As a result, the inequality ensures ${\cal V}(\gamma) = {\rm cl}({\cal V}(\gamma))$ because $u \in {\rm cl}({\cal V}(\gamma))$ implies $u \in {\cal V}(\gamma)$. Therefore, item~(ii) is proven.
\par
Finally, we verify that item~(iii) holds. We can easily see that $\bm{\bar{\lambda}} \coloneqq \frac{1}{n} \bm{1}_{n} \in \mathbb{R}^{n} = {\rm ri}(\mathbb{R}^{n})$ and $\bm{1}_{n}^{\top} \bm{\bar{\lambda}} - 1 = 0$. It then follows from item~(i) and \cite[Proposition~5.3.3]{Bertsekas2015} that the dual problem of~\eqref{eq:sofk} has at least one optimal solution, namely that item~(iii) is satisfied.  
$\square$

\end{document}